\documentclass[12pt,A4]{article}
\usepackage[top=0.65in,left=1.0in,footskip=0.75in,marginparwidth=1.0 in]{geometry}
\usepackage{cite}
\usepackage{natbib}
\usepackage{url}
\usepackage[right]{lineno}
\usepackage{eqnarray,amsmath, mathtools}
\usepackage{eqnarray,amsmath, mathtools}
\usepackage{mathrsfs}
\usepackage[utf8]{inputenc}
\usepackage{microtype}
\DisableLigatures[f]{encoding = *, family = * }
\usepackage{changepage}
\usepackage[aboveskip=3pt,labelfont=bf,labelsep=period,singlelinecheck=off]{caption}
\makeatletter
\renewcommand{\@biblabel}[1]{\quad#1.}
\makeatother
\usepackage{lastpage,fancyhdr,graphicx}
\usepackage{epstopdf}
\usepackage{rotating, hyperref}
\usepackage{color}

\definecolor{Gray}{gray}{.25}
\usepackage{graphicx}
\usepackage{sidecap}
\usepackage{wrapfig}
\usepackage[pscoord]{eso-pic}
\usepackage[fulladjust]{marginnote}
\reversemarginpar
\begin{document}

\begin{flushleft}
{\Large
{A comprehensive study on the characterization of lyzed blood samples using dual-wavelength photoacoustics}  
}
\newline
\\
Subhadip Paul\textsuperscript{1},
Hari Shankar Patel\textsuperscript{2},
Vatsala Misra\textsuperscript{3},
Ravi Rani\textsuperscript{3},
Amaresh K. Sahoo\textsuperscript{1},
Ratan K. Saha$^*,$\textsuperscript{1}
\\
\bigskip
{$^1$} \small{Department of Applied Sciences, Indian Institute of Information Technology Allahabad, Jhalwa, Prayagraj, 211015, U.P, India.}
\\
{$^2$} \small{Laser Biomedical Applications Division, Raja Ramanna Centre for Advanced Technology, Indore, 452013, M.P., India.}
\\
{$^3$} \small{Department of Pathology, Moti Lal Nehru Medical College, Prayagraj, 211002, U.P., India.}
\bigskip
\\
* ratank.saha@iiita.ac.in 
\end{flushleft}
\begin{abstract}
Anemia is a global health concern, prompting the need for rapid and accurate diagnostic tools, especially for vulnerable populations. Estimating the blood lysis level (LL) and oxygen saturation (SO$_2$) are essential not only for anemia but also for other hemolytic conditions. This study explores the potential of photoacoustic (PA) spectroscopy as a quantitative tool for evaluating hemolysis in anemia diagnosis. In vitro PA measurements on human blood samples were validated through computational modeling using the discrete dipole approximation, Monte Carlo, and k-Wave acoustic simulations. The quantitative values of blood hematocrit (H), LL and SO$_2$ have been estimated using simulated and experimental PA signals. The wavelength pairs 700-905\,nm and 700-1000\,nm have been found to be optimal for the simultaneous estimation of these parameters and provided H and SO$_2$ estimations with accuracy approximately $>90\%$ up to LL=14\% and for LL = 0-30\%, respectively. The correlation coefficient between the actual and evaluated lysis levels has been computed to be $\approx 0.90$. Further investigation is needed to enhance the robustness and clinical applicability of the developed method under an in vivo setting when both the H and LL levels are not known.
\end{abstract}
\subsubsection*{Keywords}
  Photoacoustics (PAs), Oxygen saturation level (SO$_2$), Hematocrit level (H), Lysis level (LL), Monte-Carlo simulation (MC), Discrete Dipole Approximation (DDA).
\section{Introduction}
Hemolysis is the premature destruction of red blood cells (RBCs). It occurs because of the rupture of RBC membranes and causes release of hemoglobin into the bloodstream. It is often linked with underlying health conditions \cite{Mohandas2008}. Excessive hemolysis can contribute to anemia, oxidative damage, jaundice, and renal complications, posing significant health risks \cite{Beutler2001, Kato2007, Cappellini2008, Khan2016}.
Anemia has always been a global issue, primarily because of iron deficiency. WHO defines anemia as the hemoglobin concentration falling below 12 g/dl and 13 g/dl for females and males, respectively \cite{akbarprevalence}. In a recent study by WHO, approximately 40\% of all children aged 6–59 months, 37\% of pregnant women, and 30\% of women 15–49 years of age are affected by anemia all over the globe. In southeastern countries, particularly in the Indian subcontinent, anemia cases remain exceptionally high; more than 35\% have been recorded \cite{Safiri2021}. Various factors, including autoimmune disorders, infections, genetic abnormalities, toxins, and mechanical stress, can trigger this process \cite{Zanella2014, White2018, Rees2010, Kirschner2015}. Early identification of hemolysis is crucial to mitigating complications, yet conventional diagnostic techniques rely on time-intensive laboratory analyses \cite{Shander2009}. Methods like visual inspection, Spectrometry (the most accurate one), complete blood counts (CBC) and the heptoglobin test are commonly used for the detection and quantification RBC lysis. Some complicated techniques such as Lactate Dehydrogenase, Genetic test can also be performed for the same purpose \cite{van2020simple}. 

In recent times, the photoacoustic (PA) technique has been evolving as a promising blood characterization tool. The most important feature of this technique is that, in vivo assessment of blood parameters may be possible. Saha et al. compared the PA signal properties of normal and lyzed blood samples at 532 and 1064 nm optical wavelengths \cite{saha2013photoacoustic}, \cite{saha2014probing}. Banerjee et al. studied the dynamical variation of the PA signal profile when RBCs are undergoing lysis using a low-cost 905 nm based PA device \cite{banerjee2023observing}. Bodera et al. examined blood clot and blood lysis in vitro using a high-frequency ultrasound detector-based imaging system \cite{bodera2023detection}. Besides these, researchers all over the globe are increasingly using the PA technique for probing blood samples. For example, Yadem et al. have introduced Cytophone, a PA-based flow cytometer to detect malaria infection in the human body \cite{yadem2024noninvasive}. Padmanabhan et al. have developed a PA polarization–enhanced optical rotation sensing system and found chiral molecular concentration for depths up to a few millimeters \cite{padmanabhan2025deep}. A vast amount of work has also been conducted on normal and pathological blood samples by many groups across the globe \cite{brunker2016acoustic}, \cite{esenaliev2002optoacoustic}, \cite{petrov2006multiwavelength}, \cite{petrov2017simultaneous}, \cite{karpiouk2008combined}, \cite{saha2011simulation}, \cite{saha2011effects}, \cite{saha2014computational}, \cite{hysi2012photoacoustic}, \cite{paul2024quantitative}, \cite{paul2024numerical}. \cite{biswas2017quantitative}, \cite{pai2015photoacoustics}.

PA emission from blood essentially depends upon its optical properties, namely, absorption coefficient ($\mu_{\text{abs}}$), scattering coefficient ($\mu_{\text{sca}}$) and anisotropy factor ($g$). Though these are bulk properties but they are governed by the microscopic features such as spatial organization of RBCs, morphological, biophysical and biochemical properties of individual cells \cite{saha2011simulation}, \cite{saha2011effects}, \cite{saha2014computational}, \cite{paul2024quantitative}, \cite{paul2024numerical}. There exist a number of theoretical methods and numerical schemes which can be used to obtain the optical absorption cross-section, scattering cross-section and $g$ factor of a scatterer \cite{mishchenko2000light}. For instance, the Mie theory is applicable for regular shapes \cite{wiscombe1980improved}. The Born approximation \cite{hudson1981use} and Wentzel-Kramers-Brillouin approximation work well for weak scattering or slowly varying media \cite{godin2015wentzel}. Other common approaches include the T-matrix method \cite{wang2015light}, the finite element method \cite{li2014smoothed}, the finite difference time domain method \cite{efimenko2015scattering}, the ray tracing method \cite{tan2004ray}, the Monte-Carlo method \cite{wang1995monte}, etc. One popular formulation is the discrete dipole approximation (DDA) \cite{draine1988discrete}\cite{devoe1964optical}\cite{mishchenko2000light}. It models a scatterer having an arbitrary shape as a collection of discrete electric dipoles. The knowledge of the refractive indices of both the cell and the surrounding medium are required for its implementation. It can be used to evaluate the optical parameters of RBC as well. 

Theoretical studies modeling PA emission from blood consider RBCs as spherical absorbers and thus neglect the actual shape of RBCs \cite{kaushik2021characterization}\cite{fang2009monte}\cite{saha2011simulation}. In this study, we have addressed this gap by incorporating the exact RBC morphology for numerical calculations of optical absorption cross-section, scattering cross-section and $g$ factor for a healthy RBC exploiting the DDA technique. Freely available discrete dipole scattering (DDSCAT) software implements the DDA formulation. Subsequently, obtained the optical properties ($\mu_{\text{abs}}$, $\mu_{\text{sca}}$ and $g$) of a blood sample at certain optical wavelengths. A series of lyzed blood samples with initial hematocrit, H=50\% and oxygenation, SO$_2$=68\% have been considered. The lysis level has been varied from LL=0 to 30\%. The suspending medium has been prepared by mixing hemoglobin molecules in phosphate-buffered saline (PBS) medium or in plasma (PLS) medium. DDSCAT, Monte Carlo and k-Wave simulations have been conducted sequentially to generate the PA signals \cite{paul2024numerical}. Analogous experimental studies (in vitro) have also been carried out with human blood samples. Accordingly, H, LL and SO$_2$ levels have been estimated from simulated/measured signals using dual wavelength PAs (700-905 and 700-1000 nm wavelength pairs). Simulation and experimental results reveal that accurate estimations of H and LL are possible with accuracy $>90\%$ up to LL=14\% and for SO$_2$ up to LL=30\%.  The primary contributions of this work are:
i) Determination of the optical parameters ($\mu_{\text{abs}}$, $\mu_{\text{sca}}$ and $g$) of lyzed blood samples at some specific optical wavelengths and using the DDA model incorporating the biconcave shape of RBCs. 
ii) Generation of PA signals utilizing the k-Wave toolbox based on the fluence maps provided by the Monte Carlo simulation.
iii) Measurement of PA signals from analogous human blood samples.  
iv) Quantification of the levels of H, LL and SO$_2$ from simulated and experimental PA signals. 
\section{Governing theoretical models}
%
\subsection{PA wave equation}
In practice, short laser pulses illuminate a tissue sample. Photons are absorbed by the sample, causing its rapid thermo-elastic expansion, followed by emission of PA signals. The time-dependent PA wave equation in a 3D {generalized} coordinate system can be cast as,
\begin{eqnarray}
\left( {\nabla}^2 - \frac{1}{v^2}\frac{\partial^2}{\partial t^2} \right) p(\textbf{r}, t) = -\frac{\beta}{\text{C}_P} \frac{\partial \mathscr{H}(\textbf{r}, t)}{\partial t}.
\label{PA}
\end{eqnarray}
In equation (\ref{PA}), the notations $v$, $\beta$ and C$_P$ are the speed of sound in the medium, isobaric volume expansion coefficient, and isobaric specific heat of the source region, respectively; $\mathscr{H}(\textbf{r}, t)$ the heating function, i.e., the amount of heat deposited per unit time per unit volume and $p(\textbf{r}, t)$ is the pressure at any arbitrary point at a location $\textbf{r}$ at time $t$. The initial pressure rise in the sample due to pulsed laser heating is $p_0=\mu_{\text{abs}}\Gamma F$ with $\Gamma$ and $F$ being the Gr\"uneisen parameter and fluence of the laser beam, respectively. The temporal profile of a PA signal depends upon the photon weight deposition map in the tissue. 

Photon propagation in a tissue can be modeled using the famous radiative transfer equation (RTE). The RTE is valid for both ballistic and diffusion regimes ($< 100$ $\mu m$) \cite{yao2009quantitative, saratoon2013gradient}. An approximate version of the RTE known as the diffusion equation (DE) can also model the same faithfully when $\mu_{\text{abs}} << \mu_{\text{sca}}$. The DE is given by,
\begin{multline}
\left[\hat{\zeta}.\nabla + (\mu_{\text{sca}} +\mu_{\text{abs}})\right]\mathscr{R}(\textbf{r}, \hat{\zeta})=\mu_{\text{sca}}\left( \int_{4\pi} \mathscr{R}(\textbf{r}, \hat{\zeta}) f(\hat{\zeta}.\hat{\zeta'})d\Omega'\right)+\Sigma(\textbf{r}, \hat{\zeta}),
\label{RTE3}
\end{multline}
where, $ \mathscr{R}(\textbf{r}, \hat{\zeta})$ is the time-independent radiance; $\zeta$ and $\zeta^{\prime}$ are the direction vectors of scattered and incident light; $\Sigma(\textbf{r}, \hat{\zeta})$ is the source term. If it is assumed that, $\mathscr{H}(\textbf{r}, t) = I_0 (\textbf{r})\mu_{\text{abs}} e^{-i\omega t},$ then one can write, 
\begin{eqnarray}
 \quad I_0(\textbf{r}, t)= \int_{4\pi} \tilde{\mathscr{R}}(\textbf{r}, \hat{\zeta}, t) d\Omega,
\label{H}
\end{eqnarray}
and 
\begin{eqnarray}
 \quad F(\textbf{r})= \int_{-\infty}^{\infty} I_0(\textbf{r}, t) dt,
\label{II}
\end{eqnarray}
here, $d\Omega$ indicates the solid angle through which scattered photons pass. Note that $ \tilde{\mathscr{R}}(\textbf{r}, \hat{\zeta}, t)$ is the time-dependent radiance; $ \tilde{\mathscr{R}}(\textbf{r}, \hat{\zeta}, t)$ and $\mathscr{R}(\textbf{r}, \hat{\zeta})$ are the Fourier transform pairs. Equation (\ref{RTE3}) is not analytically solvable. Thus, the Monte Carlo simulation technique is generally applied. The optical properties of the tissue ($\mu_{\text{abs}}, \mu_{\text{sca}}$ and $g$) have to be known for this purpose and thus, these quantities are defined in the next sections.
\subsection{The discrete dipole approximation}
The complete derivation of the {mathematical model of the DDA can be found else where} Draine et al. \cite{draine1988discrete, draine1993beyond, draine2008discrete}. However, the background theory has been summarized below for the sake of completeness of this study. For a fixed target geometry in a coordinate system \(\hat{x}, \hat{y}, \hat{z}\), we generate a dipole array. It aims to obtain a self-consistent set of dipole moments $\textbf{P}_j$, $ ( j = 1, \dots, N ) $ so that $\textbf{P}_j = \alpha_j \textbf{E}_{\text{loc},j}$, where, $\alpha_j$: {polarizability} and \textbf{E}$_{\text{loc}, j}$: {local} electric field at dipole point $j$. As noted by Purcell and Pennypacker (1973) and Yung (1978), {this can be, in terms of complex vectors, written as $N$ simultaneous equations of the form}\cite{purcell1973scattering}, \cite{yung1978variational}
\begin{equation}
    \textbf{P}_j = \alpha_j \left( \textbf{E}_{\text{inc},j} - \sum_{m \neq j} \textbf{A}_{jm} \textbf{P}_m \right),
    \label{eq:dipole_eqn}
\end{equation}
 $\textbf{E}_{\text{inc},j}$ is the electric field at position $j$ due to the incident plane wave,
   $ \textbf{E}_{\text{inc},j} = \textbf{E}_0 \exp (i\textbf{k} \cdot \textbf{r}_j - i \omega t),$
and $\textbf{A}_{jm} \textbf{P}_m$ is the contribution to the electric field at position $j$ due to the dipole at position $m$:
\begin{equation}
\begin{aligned}\label{AP}
    \textbf{A}_{jm} \textbf{P}_m &= \frac{\exp(ikr_{jm})}{r_{jm}^3} 
    \Bigg\{ k^2 \textbf{r}_{jm} \times (\textbf{r}_{jm} \times \textbf{P}_m)  + \frac{(1 - ikr_{jm})}{r_{jm}^2}
    \Big[\textbf{r}_{jm}^2 \textbf{P}_m - 3\textbf{r}_{jm} (\textbf{r}_{jm} \cdot\textbf{P}_m) \Big] \Bigg\}, \\ \quad (j \neq m).
\end{aligned}
\end{equation}
where $r_{jm} = |\textbf{r}_j - \textbf{r}_m|$. Equation (\ref{AP}) serves to define the matrices $\textbf{A}_{jm}$ for $ j\neq m$. It is convenient to define also matrices $ \textbf{A}_{jj} = \alpha_j^{-1}$,
so that the scattering problem can be compactly formulated as a set of $N$ inhomogeneous linear complex vector equations:
\begin{equation}
    \sum_{m=1}^{N} \textbf{A}_{jm} \textbf{P}_m = \textbf{E}_{\text{inc},j} \quad (j = 1, \dots, N),
\end{equation}
here, $(A_{jl})_{mi} = (A_{lj})_{im}$, i.e, the $\textbf{A}_{jm}$ are $3 \times 3$ matrices symmetric.
\subsection{Extinction, absorption and scattering efficiency factors}
Once the polarisation $\textbf{P}_j$ is known, the extinction coefficient factor for the grain is computed from the forward-scattering amplitude using the optical theorem,\cite{Schiff1968Quantum,draine2008discrete}
\begin{equation}
    \text{Q}_{\text{ext}}=\frac{4 k}{\mathtt{a^2_{eff}}|\mathbf{E}_{inc}|^2}\sum^N_{j=1} \text{Im}(\mathbf{E}^*_{inc, j}\cdot \mathbf{P}_j),
    \label{optical_theorem1}
\end{equation}
$\mathtt{a_{eff}}$ is the effective radius of a RBC. The absorption coefficient factor is obtained by summing over the energy dissipation rate of each dipole. By substituting $\mathbf{P} = \alpha \mathbf{E}_{inc}$ into Eq. (\ref{optical_theorem1}), the extinction coefficient factor can readily be obtained from the optical theorem as, \cite{draine1988discrete}
\begin{eqnarray}
    \text{Q}_{\text{ext}}=\frac{4 k}{\mathtt{a^2_{eff}}|\mathbf{E}_{\text{inc}}|^2}\sum^N_{j=1}  \text{Im}(\mathbf{P}\cdot({\alpha^{-1}})^* \mathbf{P}^*).
    \label{optical_theorem2}
\end{eqnarray}
Note that for isotropic polarizability it reduces to $\text{Q}_{\text{ext}}=4\pi k  \text{Im} (\alpha)$.

Thus, the absorption coefficient factor for the entire grain is \cite{draine1994discrete}
\begin{equation}
    \text{Q}_{\text{abs}}=\frac{4 k}{\mathtt{a^2_{eff}}|\mathbf{E}_{\mbox{inc}}|^2}\sum^N_{j=1} \left\{   \text{Im} [\mathbf{P_j}\cdot({\alpha_j^{-1}})^* \mathbf{P_j}^*] - \frac{2 k^3}{3} \mathbf{P_j}\cdot{\mathbf{P_j}}^*   \right\}.
    \label{DDAQabs}
\end{equation}
The scattering coefficient factor can, in principle, be obtained from the difference between the extinction and absorption coefficient factors: \cite{draine1994discrete}
\begin{eqnarray}
    \text{Q}_{\text{sca}} = \frac{k^4}{\pi \mathtt{a^2_{eff}} |\mathbf{E}_{\text{inc}}|^2} 
    \int d\Omega \Bigg| \sum_{j=1}^{N} \left[ \textbf{P}_j - \hat{n}(\hat{n} \cdot \textbf{P}_j) \right] 
    \exp{(-ik\hat{n} \cdot \textbf{r}_j)}\Bigg|^2,
    \label{DDAQsca}
\end{eqnarray}
where \( \hat{n} \) is a unit vector in the direction of scattering. It is also of interest to evaluate the scattering asymmetry parameter, $g \equiv \langle \cos \theta \rangle$, which is given by, \cite{draine1988discrete}
\begin{eqnarray}
    g  =
    \frac{k^3}{\pi \mathtt{a^2_{eff}}\text{Q}_{\text{sca}} |\mathbf{E}_{\text{inc}}|^2} 
    \int d\Omega \, (\hat{n} \cdot \mathbf{k}) 
    \Bigg| \sum_{j=1}^{N} 
    \left[ \textbf{P}_j - \hat{n}(\hat{n} \cdot \textbf{P}_j) \right] 
    \nonumber 
    \exp{(-ik\hat{n} \cdot \textbf{r}_j)} \Bigg|^2.
    \label{DDAg}
\end{eqnarray}
It varies from -1 to 1.
%
%
\begin{figure}[!t]
    \centering
\includegraphics[width=0.7\linewidth]{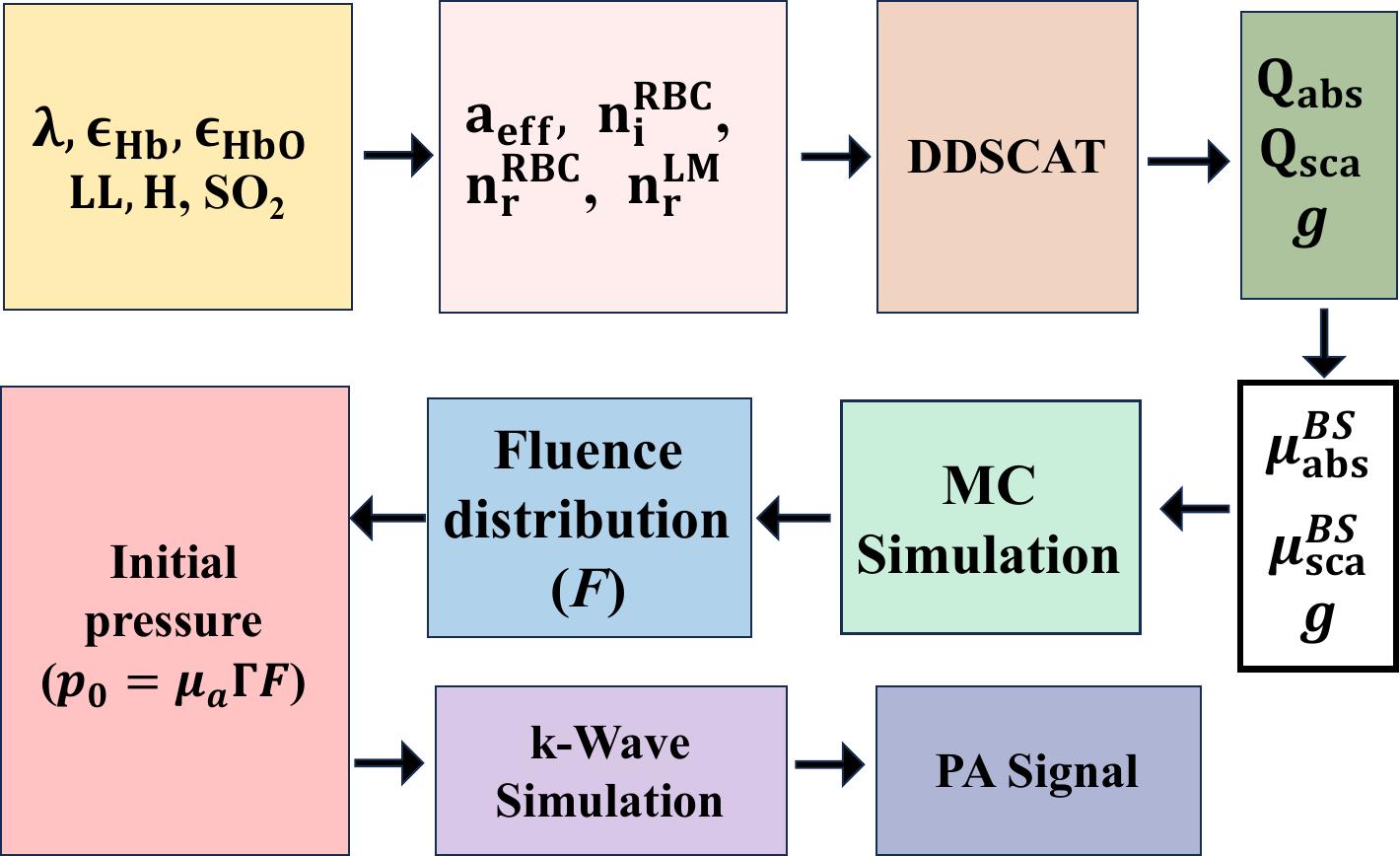}
    \caption{Workflow diagram illustrating the simulation pipeline implemented for modeling various lysis levels of RBC samples in PBS and PLS suspending media.}
    \label{flowchart1}
\end{figure}
\begin{table}[!b]
    \centering
        \caption{Tabulated values of the real ($n_r^{\mbox{LM}}$) and imaginary ($n_i^{\mbox{LM}}$) components of the refractive index for the suspending lyzed media (PBS/PLS based) across different stages of lysis.}
\label{nk_rbc_pbs_scott_Med}
\includegraphics[width=0.8\linewidth]{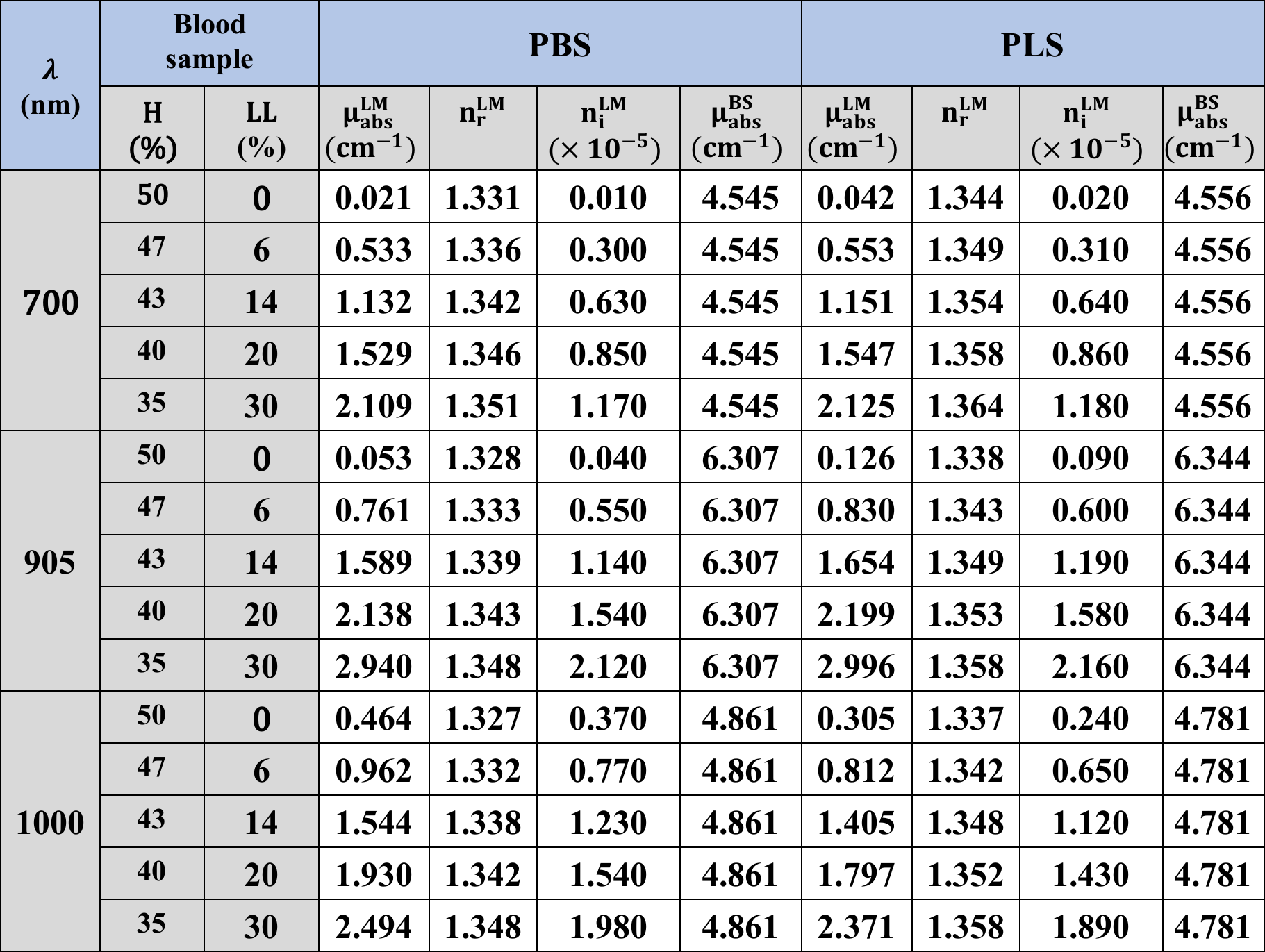}
\end{table}
\section{Materials and methods}
\subsection{Numerical investigation}
\subsubsection{Calculation of optical absorption coefficient and complex refractive index}
A healthy RBC typically occupies a volume of $V_{\mbox{RBC}}=91.52$ fL and retains 281 million hemoglobin molecules having a molar mass of 64500 g; 66\% of $V_{\mbox{RBC}}$ is water \cite{paul2024numerical, shapovalov2020geometry}. Based on these, the optical absorption coefficient for the semi-solid medium present inside RBC could be expressed as,
\begin{eqnarray}
\mu_{\text{abs}}^{\text{RBC}} = 2.303 \left( \text{C}_{\text{HbO}}   \varepsilon_{\text{HbO}} + 
\text{C}_{\text{Hb}}   \varepsilon_{\text{Hb}} \right)+ 0.66   \mu_{\text{abs}}^{\text{W}},
\label{single_cell_mua}
\end{eqnarray}
where $\text{C}_{\text{HbO}}$ and $ \varepsilon_{\text{HbO}}$ are the molar concentration and extinction coefficient for oxy-hemoglobin (HbO), respectively; the same quantities for deoxy-hemoglobin (Hb) are denoted by $\text{C}_{\text{Hb}}$ and $ \varepsilon_{\text{Hb}}$, respectively; $\mu_{\text{abs}}^{\text{W}}$ is the optical absorption coefficient for water (\url{https://www.omlc.in})
. Accordingly, real and imaginary parts of the refractive index for the same cellular matrix could be estimated to be \cite{fribel2006determination},
\begin{eqnarray}
\mathtt{n_r^{\text{RBC}}} &=& \mathtt{n_r^{\text{M}}} [\mathscr{B} (\text{C}_{\text{HbO}}+\text{C}_{\text{Hb}}) + 1], 
\\ 
\mathtt{n_i^{\text{RBC}}} &=& \frac{\mathtt{\mu_{\text{abs}}^{\text{RBC}}}  \lambda}{ 4 \pi},
\label{refractive_index-CM}
\end{eqnarray}
respectively. Here, $\mathscr{B}$: wavelength-dependent specific refractive index increment \cite{li2000refractive}. The numerical values of these quantities for a RBC with oxygen saturation $\text{SO}_2=0.68$ could be computed to be, $\mu_{\text{abs}}^{\text{rbc}}=9.070, 12.563 \mbox{ and } 9.258$ cm$^{-1}$, $\mathtt{n_r^{\text{rbc}}}=1.418, 1.415 \mbox{ and } 1.417$ and $\mathtt{n_i^{\text{rbc}}}=5.050\times 10^{-5}, 9.050\times 10^{-5} \mbox{ and } 7.370\times 10^{-5}$ at 700, 905 and 1000 nm, respectively, assuming $\text{C}_{\text{HbO}}+\text{C}_{\text{Hb}}=5.1\times 10^{-3}$ mole/L.

The optical absorption coefficient for the extracellular matrix significantly changes in the presence of freely suspending hemoglobin molecules, occurring because of the lysis of RBCs. It could be estimated by evaluating the following formula,
\begin{eqnarray}
\mu_{\text{abs}}^{\text{LM}} = \frac{1}{(1-\text{H}+\text{H LL})}\Big[ \text{H LL} \mu_{\text{abs}}^{\text{RBC}} +
(1-\text{H})\mu_{\text{abs}}^{\text{M}}\Big],
\label{lyzed_blood_medium}
\end{eqnarray}
where, $\mbox{LL}$ is the lysis level and $\mbox{H}$ is the hematocrit (at $\mbox{LL}=0\%$) of the sample; the subscript LM indicates PBS or PLS based lyzed medium. Table \ref{nk_rbc_pbs_scott_Med} displays the computed values of $\mu_{\text{abs}}^{\text{LM}}$ and refractive index for a series of blood samples having different lysis levels but with $\mbox{H}=0.50$ and $\text{SO}_2=0.68$. The same quantity for the whole blood ($\mu_{\text{abs}}^{\text{BS}}$) could also be determined as (by adding the contributions from intact RBCs and surrounding medium),
\begin{eqnarray}
\mu_{\text{abs}}^{\text{BS}} = \text{H}\big(1-\text{LL} \big)\mu_{\text{abs}}^{\text{RBC}} + \big( 1-\text{H}+\text{H LL}  \big)\mu_{\text{abs}}^{\text{LM}},
\label{lyzed_blood_sample_Mua}
\end{eqnarray}
subscript BS states blood sample; its numerical values are included in the same table (columns 6 and 10 for PBS and PLS media, respectively of Table \ref{nk_rbc_pbs_scott_Med}).
%
%
%
\subsubsection{DDA simulation}
\begin{figure}[!t]
    \centering
    \includegraphics[width=0.6\linewidth]{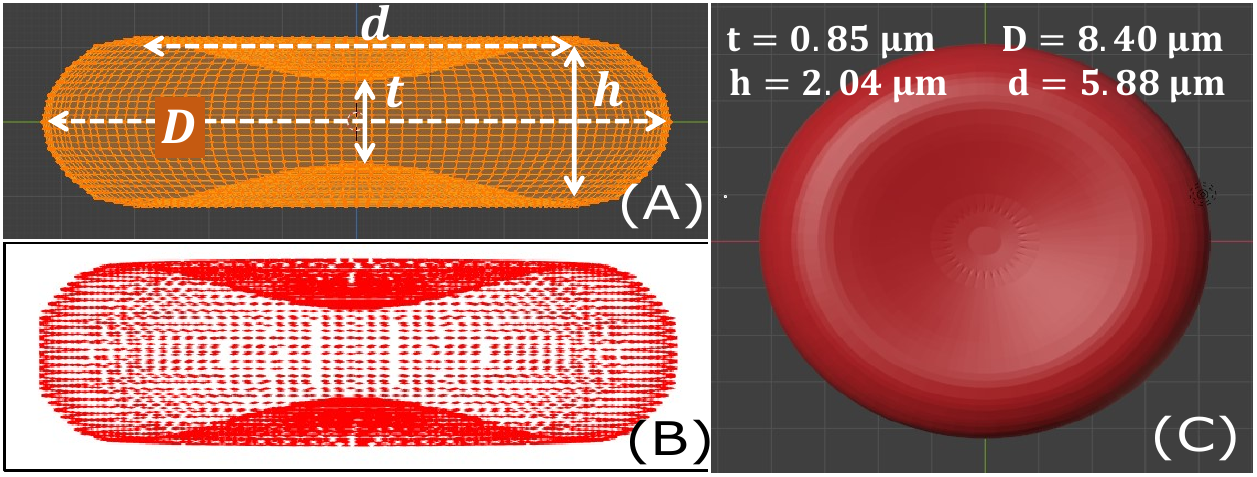}
    \caption{Visualization of red blood cell (RBC) geometry. (A) Side view of the RBC structure. (B) Dipole distribution used for electromagnetic modeling. (C) Top view highlighting the characteristic biconcave shape.}
    \label{obj_rbc}
\end{figure}

The RBC model was designed in Blender 4.2, an open-source software. Normal RBC looks like a biconcave disk \cite{kaushik2021characterization, bi2013modeling, shapovalov2020geometry}. Figure \ref{obj_rbc} shows the meridional cross-section of a biconcave RBC, defined by four critical morphological parameters: the diameter $(D)$, the dimple thickness $(t)$, the maximum thickness $(h)$, and the diameter $(d)$ of the circle determining the location of the maximum thickness. The chosen values of that morphological parameter were $D=8.40$ $\mu$m, $t=0.85$ $\mu$m, $h=2.04$ $\mu$m, and $d=5.88$ $\mu$m providing an effective radius, $\mathtt{a_{eff}}=2.79$ $\mu$m. 

The DDSCAT module is an open-source Fortran-based algorithm \cite{draine1994discrete}.
The DDSCAT version 7.3.1 was employed, and single precision mode was used to execute the DDA simulations.
The DDSCAT simulation was configured by selecting specific options for key parameters. Radiative torque calculations were disabled. The conjugate gradient solver was chosen as preconditioned Bi-Conjugate Gradient Stabilized. The Fast Fourier Transforms were performed using the Generalized Prime Factor Algorithm method. The dipole polarizability was computed using the Generalized Kernel Dipole Layered Dipole Regularization technique. Finally, binary output files were avoided during the simulation. The computational volume in DDSCAT was set to $230 \times 230 \times 80$ voxels containing cubic lattice points. The RBC model was discretized into approximately $1.6\times 10^6$ dipoles for the wavelengths of interest (i.e., 700, 905, 1000 nm) using the DDSCAT-convert module available on \url{nanoHUB.org} [see Fig. \ref{obj_rbc}(B)]. The refractive indices ($\mathtt{n_r}$ and $\mathtt{n_i}$) of the RBC and the surrounding medium were determined and {presented in} Table \ref{nk_rbc_pbs_scott_Med}). The error tolerance and maximum number of iterations were set to $10^{-5}$ and $1000$, respectively. A total of 125 orientation-average was considered to yield $\text{Q}_{\text{abs}}, \text{Q}_{\text{sca}}$ and $g$. {All computations were performed on a high-performance workstation equipped with 256 GB RAM, an AMD Ryzen Threadripper PRO 5965WX 24-core processor (3.80 GHz), 932 GB SSD, and 7.28 TB HDD storage}. Depending on the chosen $\text{H}$ and $\text{LL}$ levels, $\mu_{\text{abs}}^{\text{BS}}$ and $\mu_{\text{sca}}^{\text{BS}}$ were calculated using the formulae given below \cite{bosschaart2014literature}, \cite{paul2024quantitative}, \cite{paul2024numerical}
\begin{align} \label{muamus_dda}
    \mathtt{\mu_{abs}^{BS}} &= \frac{\text{H}}V_{\text{RBC}} \text{Q}_{\text{abs}} \pi \mathtt{a^2_{eff}}, \\
    \mathtt{\mu_{sca}^{BS}} &= \frac{\text{H}}V_{\text{RBC}} (1 - \text{H})^2 \text{Q}_{\text{sca}} \pi \mathtt{a^2_{eff}},
\end{align}
{numerical value of $g$ was provided by the DDSCAT as well}.  
\subsubsection{Monte-Carlo and k-Wave simulations}
The Monte Carlo Multilayered (MCML) algorithm was utilized to model photon transport in homogeneous blood samples vis-à-vis to obtain the fluence matrix ($F$) \cite{wang1995monte}. The simulation domain consisted of a cubic tissue volume with dimensions $200\times200\times200$ voxels, where each voxel was fixed to be $dx \times dy \times dz = 0.005 \times 0.005 \times 0.005$ cm$^3$. The optical properties ($\mu_{\text{abs}}^{\text{BS}}$, $\mu_{\text{sca}}^{\text{BS}}$, and $g$) at a particular wavelength of each voxel were assigned according to Table \ref{DDSCATopticalPRAM_pbs}, corresponding to the distinct blood sample under investigation. We introduced $\pm1\%$ random fluctuations in $\mu_{\text{abs}}^{\text{BS}}$ for 10\% of randomly selected voxels to introduce a little randomness, simulating a real condition. The incident laser beam had a diameter of $D_B = 0.10$ cm and featured a uniform lateral spatial profile along with a delta-function temporal profile. For each simulation, $2 \times 10^6$ photons were launched with reflections enabled only at the top surface ($z = 0$), while for the side and bottom walls, reflections were disabled. The fluence maps were generated for a specific sample at $\lambda = 700$, 905, and 1000 nm. A schematic of the simulation setup was shown in Fig. 3 of \cite{paul2024quantitative}

After calculating the fluence matrix ($F$), the k-Wave toolbox was used to perform 3D acoustic {simulations} (Fig.\ref{flowchart1}). The initial pressure rise was computed as $p_0 =
\Gamma \mu_{\text{abs}} F$, where $\Gamma = 1$ for all samples. The simulation domain size was $240\times240\times280$ with a grid spacing of $dx = dy = dz = 0.005$ cm. A perfectly matched layer (PML) of thickness 0.05 cm was employed, with an anisotropic absorption coefficient of 100 np/m to diminish boundary reflections. A spherically focused circular aperture ultrasound
transducer (UST) with a radius of 0.05 cm was placed at ($120, 120, 260$) grid location. Focal point of the transducer was fixed at the center of the computational domain ($120, 120, 100$). The sensor consisted of 317 grid points, with a center frequency of 5 MHz and a 70\% fractional bandwidth. The speed of sound ($v_s = v_{M} = 1500$~m/s)
and density ($\rho_s = \rho_{M} = 1000$~kg/m$^3$) remained uniform throughout the computational domain. The Courant–Friedrichs–Lewy number was set to 0.3, achieving numerical stability and providing a time step of 20 ns. A Gaussian noise was added to the simulated pressure signals to achieve 40~dB signal-to-noise ratio.
\begin{table}[!t]
    \centering
   \caption{DDSCAT based computation of absorption, scattering, and anisotropy coefficients across different lysis levels in an ambient environment (initial surrounding medium was PBS or PLS).}
    \label{DDSCATopticalPRAM_pbs}
    \includegraphics[width=0.8\linewidth]{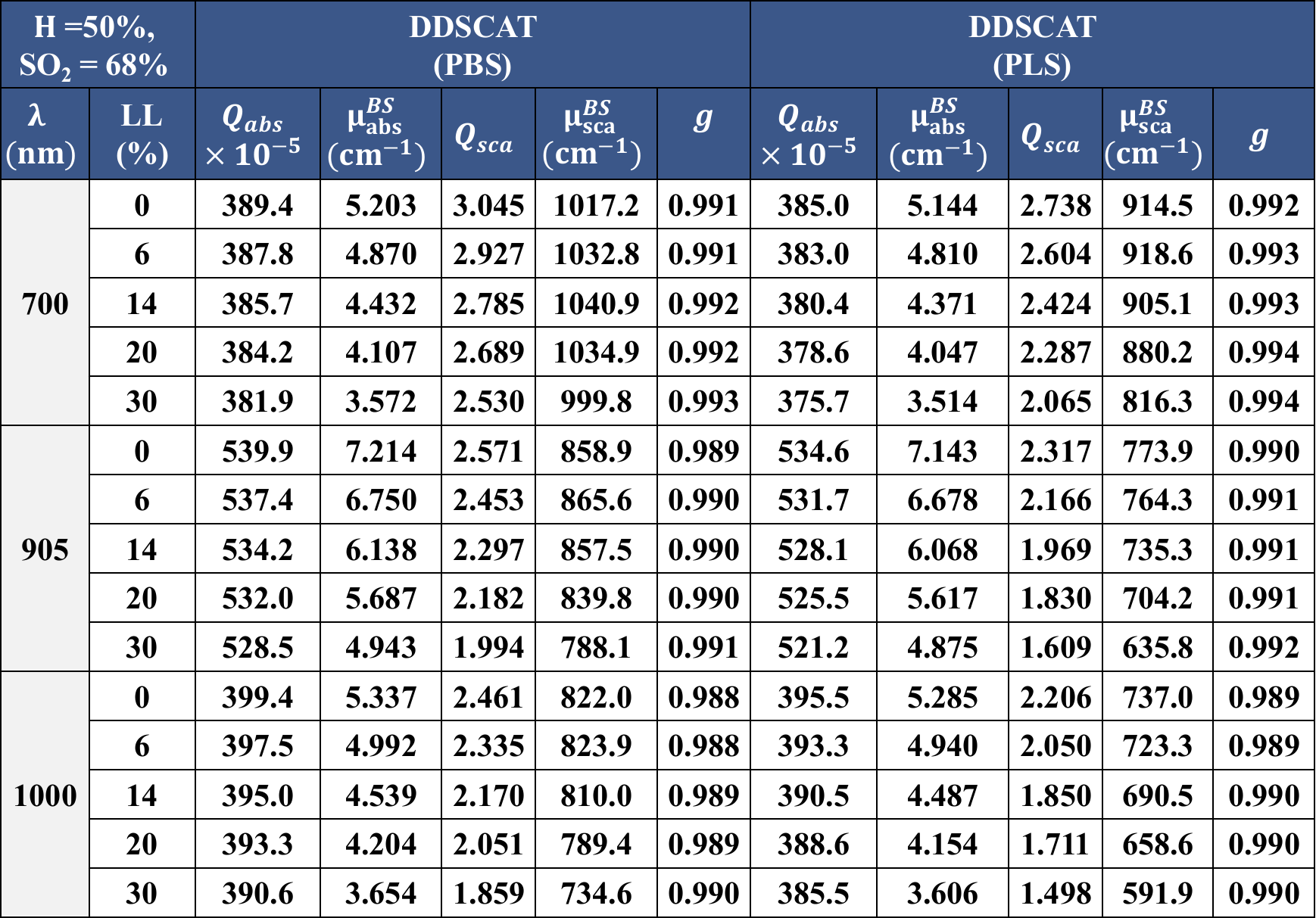}
\end{table}
\subsection{Experimental procedure}
\subsubsection{Sample preparation}
\begin{figure}[!b]
    \centering
    \includegraphics[width=0.9\linewidth]{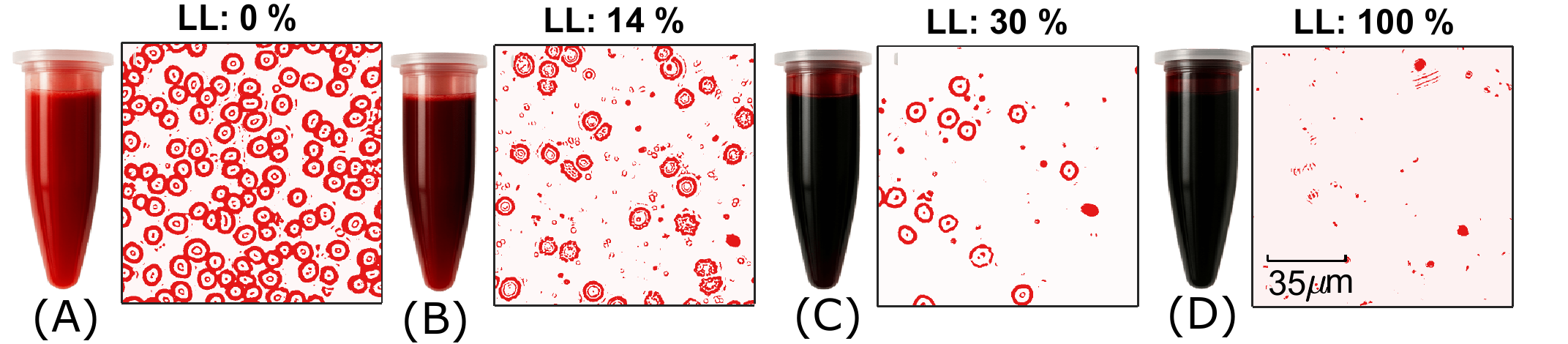}
    \caption{The figure represents (A–D) Blood samples at the lysis levels of LL=0\%, 14\%, 30\%, and 100\%, with corresponding microscopic images of smeared samples on glass slides. Scale bar: 35 $\mu$m.}
    \label{microscopy}
\end{figure}
Fresh human whole blood samples were obtained from {a} local blood bank. All five donors were healthy individuals aged between 25 and 40 years. The samples were collected in vacutainer tubes containing EDTA as an anticoagulant agent. {A complete report of various blood parameters} for each donor is provided in Table \ref{donor}. This project was cleared apriori by the ethical committee of the institute.
\begin{table}[!b]
    \centering
        \caption{Summary of demographic details and hematological indices of blood donors, showing inter-individual variability in key parameters including hematocrit (H), total hemoglobin (THB), mean corpuscular volume (MCV), mean corpuscular hemoglobin (MCH), mean corpuscular hemoglobin concentration (MCHC), red cell distribution width (RDW), and red blood cell count (RBC).}
    \label{donor}
\includegraphics[width=0.85\linewidth]{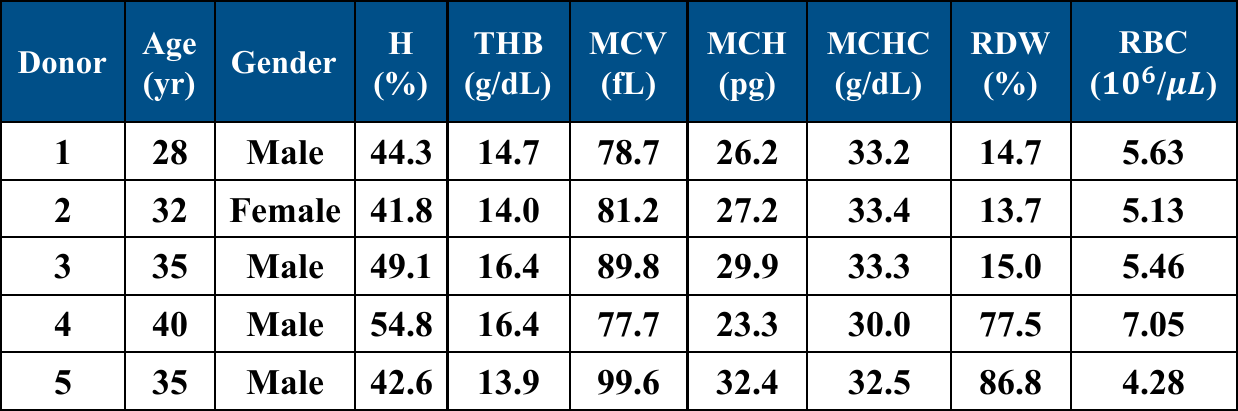}
\end{table}
For sample preparation, the blood samples were centrifuged at  ×1000 g for 15 minutes at room temperature. Blood plasma was carefully extracted and the buffy coat layer was also meticulously discarded. The packed RBCs were then evenly divided into two separate tubes. Plasma was poured into one tube, while phosphate-buffered saline (PBS) was added to the other, ensuring that both samples maintained the same hematocrit level of 50\%. This procedure was consistently followed for all donors to prepare a total of 10 samples. In the next step, individual samples were further divided into two parts. One part was subject to complete lysis using an ultrasound sonicator (Qsonica, XL-2000 series, 5W emission, 5-second pulse). The lyzed solution was further centrifuged to discard the debris. The other half was left intact.  A series of lyzed samples was prepared by mixing the 100\% and 0\% lyzed solutions in different ratios, achieving $\text{LL}=0$\%, 6\%, 14\%, 20\%, and 30\% lysis levels. A blood sample of approximately 800 µL was placed in a cylindrical sample holder for subsequent PA experiment. Some representative images of lyzed blood samples and the corresponding microscopic images are demonstrated in Fig. \ref{microscopy}. The number of cells decreases with increasing lysis. 
\begin{figure}[!t]
    \centering
    \includegraphics[width=0.85\linewidth]{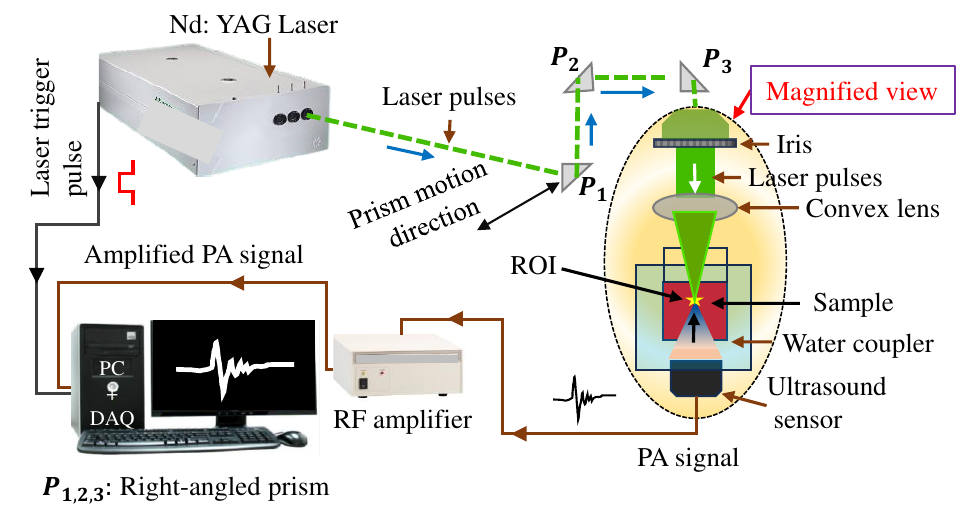}
    \caption{Schematic diagram of {the} Nd:YAG laser-based PA experimental setup.}
    \label{Experimntal_setup1}
\end{figure}
\subsubsection{PA signal detection}
The PA experimental setup is shown in Fig. \ref{Experimntal_setup1}.  A Nd: YAG laser source (EKSPLA NT352) was employed to perform the measurements. It is a solid-state, Q-switched, tunable and pulsed laser source emitting pulses of 6 ns width at a 10 Hz repetition frequency. This instrument is capable of emitting a range of wavelengths from 335 nm to 2500 nm (excluding the optical range 501-659 nm, but including 532 nm). Three prisms ($P_1, P_2, P_3$) were used to guide the laser beam. The beam was shaped and focused to a spot using an iris and a convex lens. A spherically focused ultrasound transducer (ISR053) with a center frequency of 5 MHz, bandwidth of 70\% and focal length of 25 mm was mounted at the bottom of the sample holder. It was acoustically coupled through water. A data acquisition card (ADLINK PCIe-9852) was used to record the captured PA signals at a sampling frequency of $50$ MHz; each signal contained 5000 samples. For each blood sample, a total of 100 RF signals were stored. The PA measurements were conducted at optical wavelengths of 700, 905, and 1000 nm, using optical fluences of 5.5, 19.0, and 47 mJ/cm$^2$, respectively, to remain within the ANSI safety limits \cite{ANSI-2007}. The exact H level was determined using a micro-capillary centrifuge rotating at 8500 rpm for 2 minutes. UV-Vis spectrophotometric data were also collected for the blood samples with 0\% and 30\% lysis levels between the wavelength range 500-1100 nm (PerkinElmer Lambda 365+). The blood SO$_2$ level was assessed {utilizing} measured UV-Vis data.
\subsection{Analysis of PA signals and quantification of Hb concentration and SO$_2$ levels}
The total hemoglobin concentration (THB) and oxygen saturation of a blood sample can be assessed if PA signals are measured at two optical wavelengths by exploiting the following formulae  \cite{hysi2012photoacoustic, saha2011effects},
\begin{equation}
\textrm{THB} = \frac{P_p(\lambda_1) \Delta\epsilon^{(\lambda_2)} -P_p(\lambda_2) \Delta\epsilon^{(\lambda_1)}}{\epsilon^{(\lambda_1)}_{Hb}\epsilon^{(\lambda_2)}_{HbO} -\epsilon^{(\lambda_2)}_{Hb}\epsilon^{(\lambda_1)}_{HbO}}
{\label{THB1}}
\end{equation}
and
\begin{eqnarray}
\text{SO}_2 = \frac{P_p(\lambda_2)\epsilon_{Hb}^{(\lambda_1)} - P_p(\lambda_1) \epsilon_{Hb}^{(\lambda_2)}}{P_p(\lambda_1) \Delta\epsilon^{(\lambda_2)}_{b} - P_p(\lambda_2)\Delta\epsilon^{(\lambda_1)}_{b}}
{\label{so21}}
\end{eqnarray}  
respectively and $ \Delta\epsilon^{(\lambda)}_{b}= \epsilon_{\text{HbO}}^{(\lambda)}-\epsilon_{\text{Hb}}^{(\lambda)}$. Here, $P_p$ is the peak-to-peak amplitude of the PA signal. The blood sample with highest hematocrit for each batch was considered as the calibration sample using which the H levels were determined for other samples.
\section{Results}
\subsection{Simulation results}
\begin{figure}[!t]
    \centering
    \includegraphics[width=0.8\linewidth]{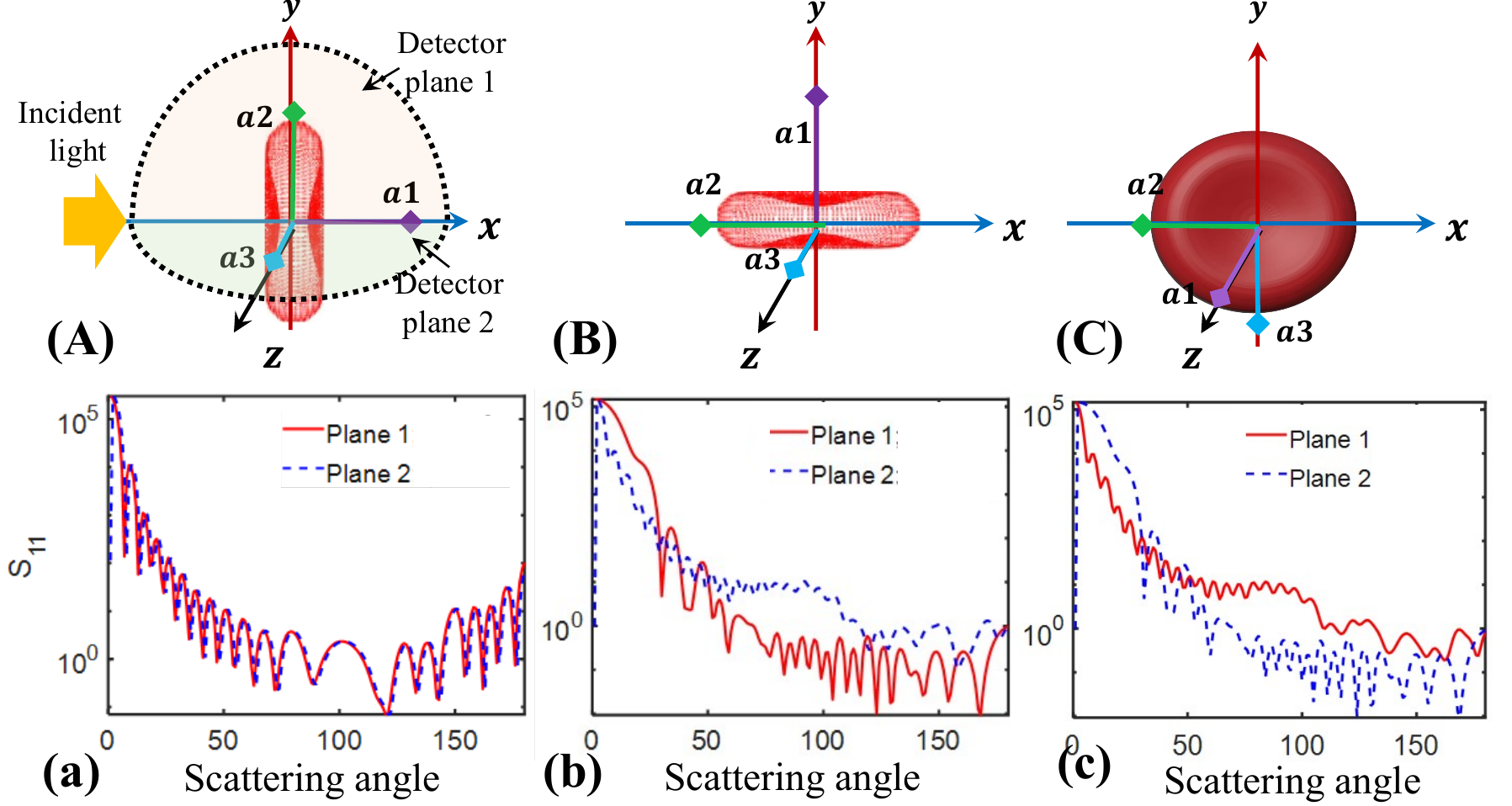}
    \caption{(A)-(C) The schematic diagram showing various orientations of RBC; $a1$, $a2$ and $a3$ are the axes associated with the object frame. The direction of the incident light and detection planes are defined with respect to the laboratory frame, in (A). (a)-(c) The corresponding plots of angular distribution of scattering amplitude ($S_{11}$) computed at two specific scattering planes for 1000 nm incident optical wavelength.}
    \label{RBCorientation}
\end{figure}

\begin{figure}[!h]
    \centering
    \includegraphics[width=0.8\linewidth]{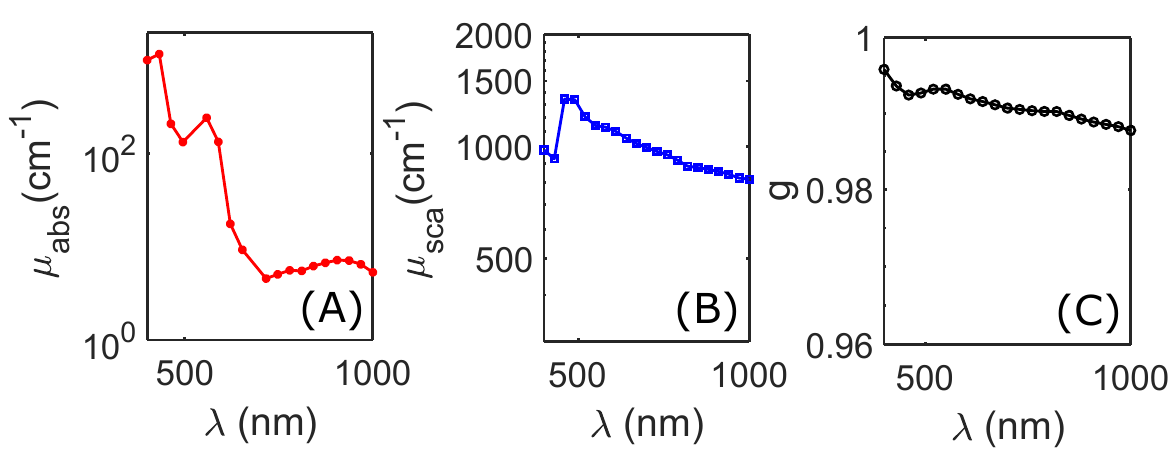}
    \caption{The plots of- (A) absorption coefficient ($\mu_\text{abs}$), (B) scattering coefficient ($\mu_\text{sca}$) and (C) anisotropy coefficient ($g$) spectra for a blood sample with hematocrit H = 50\%, $\text{SO}_2 = 68\%$ and $\text{LL}=0\%$. The results are generated using the DDSCAT module with PBS as the surrounding medium.}
    \label{optical_param}
\end{figure}

Let us begin with angular scattering patterns for a healthy RBC generated by the DDSCAT module. We considered three orientations of the scattering object as can be visualized from Fig. \ref{RBCorientation}(A)-(C). The direction of the incident light is shown in Fig. \ref{RBCorientation}(A). The scattering planes are chosen perpendicular to each other, i.e., lying on the xy and zx-planes, and the detectors are placed over 0-180$^o$. Figure \ref{RBCorientation}(a)-(c) plots the corresponding angular distributions of scattering amplitude ($S_{11}$) for 1000 nm incident light recorded at those scattering planes. It is evident from Fig. \ref{RBCorientation}(a) that the calculated $S_{11}$ for both the detection planes are overlapped since the source look identical for these planes. Similarly, Fig. \ref{RBCorientation}(b) and (c) presents that $S_{11}$ curves are getting separated and switch position because of the symmetry possessed by the biconcave shape. The $\mu_{\text{abs}}$,  $\mu_{\text{sca}}$, and $g$ spectra were calculated using the DDSCAT software are plotted in Fig. \ref{optical_param}(A)-(C) for an optical spectral range of 400-1000 nm with H $=$50\%, $\text{SO}_2 = 68\%$, LL $=$ 0\% and for the PBS suspending medium.


\begin{figure}[!t]
    \centering
\includegraphics[width=0.8\linewidth]{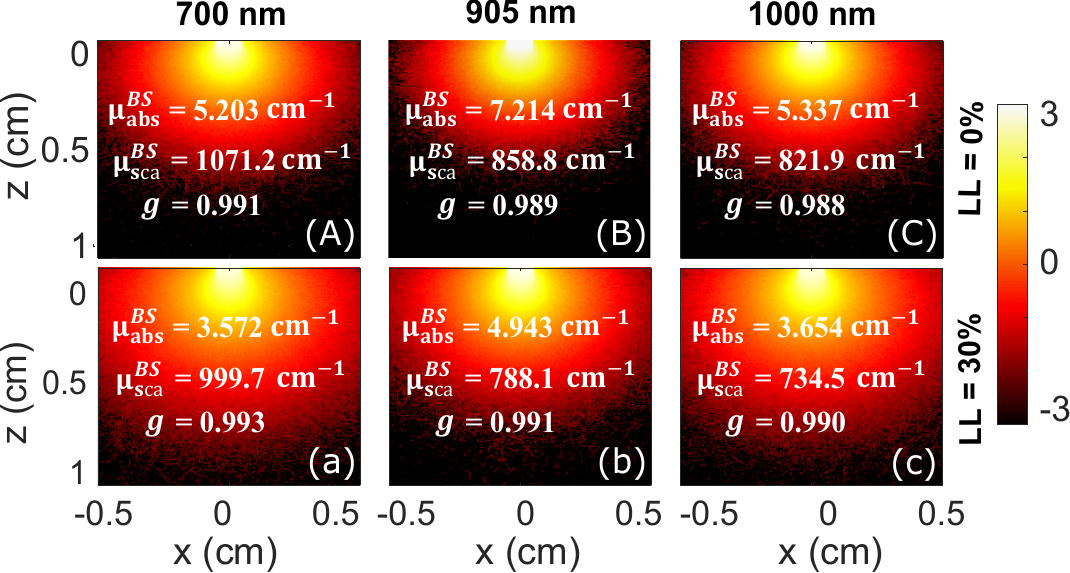}
    \caption{Cross-sectional views of fluence distribution in PBS-based lyzed blood samples at 700, 905 and 1000 nm for LL=0\% and 30\% lysis levels with SO$_2$ = 68\% and initial hematocrit level, H=50\%.}
    \label{McFluence}
\end{figure}

\begin{figure}[!b]
    \centering
\includegraphics[width=0.8\linewidth]{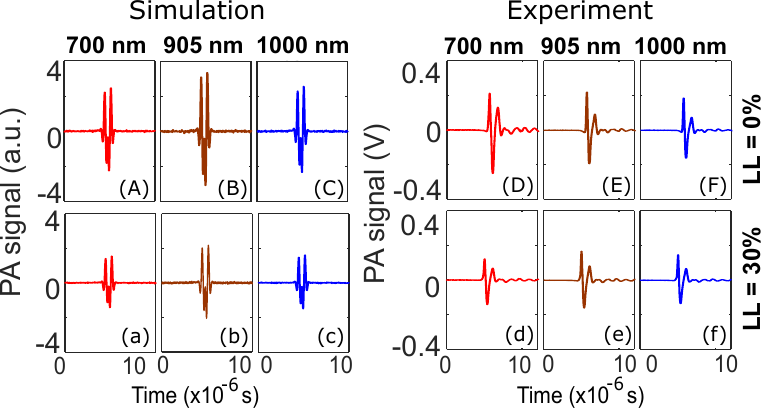}
    \caption{Left panel: Demonstration of the simulated PA signals for LL=0\% (A-C, top panel) and LL=30\% (a-c, bottom panel) lyzed blood samples at 700, 905, and 1000 nm wavelengths, SO$_2$ = 68\% and initial hematocrit level, H=50\%. Right panel: Similar plots for the experimentally measured PA signals for the sample collected from donor 1.}
    \label{PAsignalSim}
\end{figure}

\begin{figure}[!h]
    \centering
        {
        \includegraphics[width=0.8\textheight]{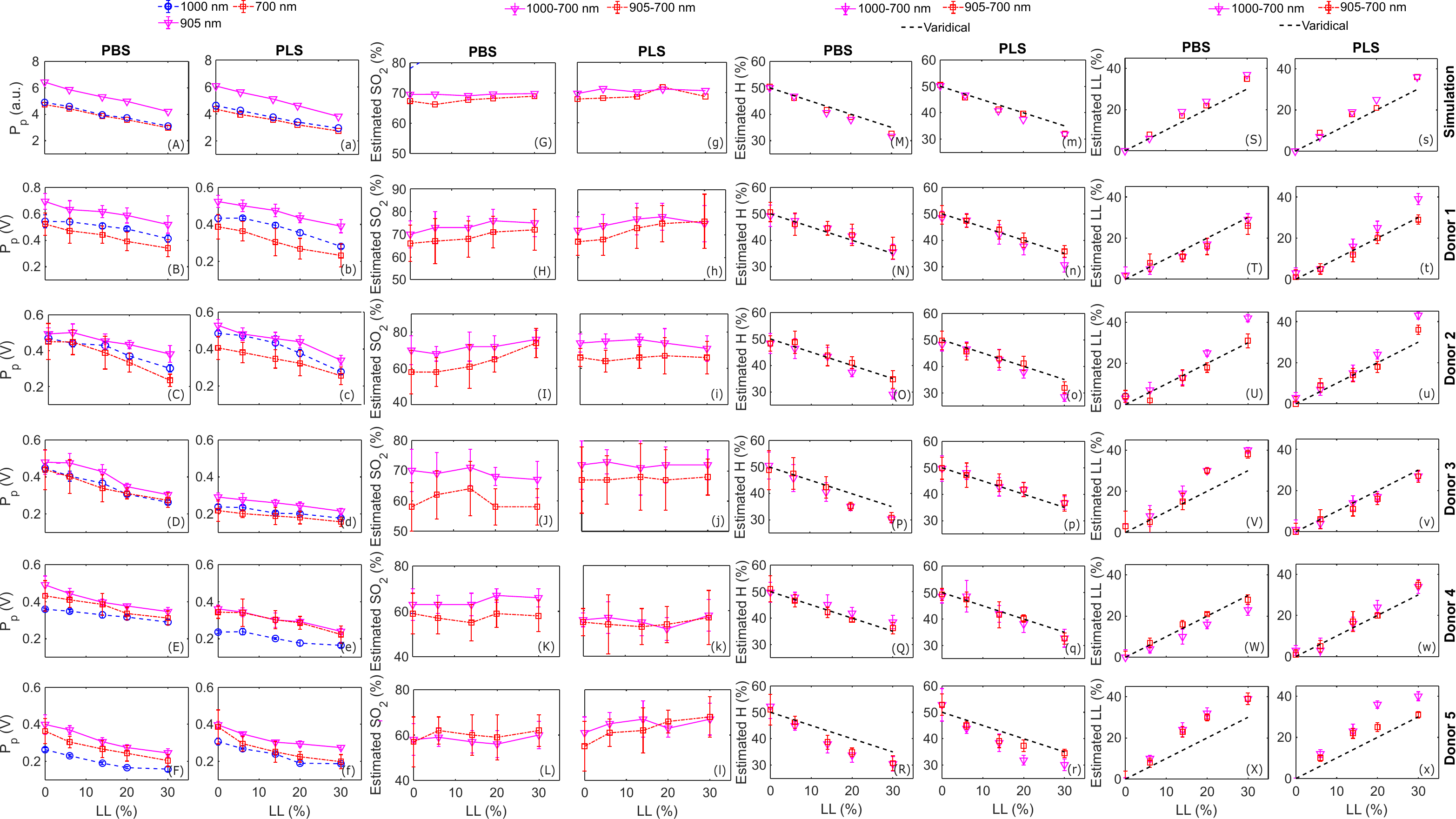}
    }
\caption{Top row: Analysis of simulated PA signals for the assessment of blood oxygenation, hematocrit and lysis levels. Rows 2-6: Analysis of experimentally measured PA signals for five different donors (see Fig. \ref{donor} for details) and quantification of the same parameters (mean$\pm$std) using the dual wavelength protocol.} 
\label{fig:SimD1_5_exp}
\end{figure}

Photon-weight deposition maps generated by the MC simulation are shown in Fig. \ref{McFluence} for blood samples with initial hematocrit level, H=50\% and lysis levels LL=0\% and 30\% for the illuminating wavelengths of 700, 905, and 1000 nm. The nominal oxygenation level is fixed at SO$_2$ = 68\% and PBS is the suspending medium. A color bar tied to the figure quantifies the color code. The optical parameters ($\mu_{\text{abs}}^{\text{BS}}$, $\mu_{\text{sca}}^{\text{BS}}$, and $g$) are provided in each figure. The photons incident from the top surface, $\text{z}=0$. It can be seen from this figure that photons are diffused vis-\'{a}-vis penetrate less in the blood sample with LL=0\% compared to that of LL=30\% [compare top and bottom rows of Fig. \ref{McFluence}]. This is  because $\mu_{\text{abs}}^{\text{BS}}$ is higher at the former sample than the later one and is also true for all incident optical wavelengths.
 In both media, $P_p$ values at 905 nm are higher than those at 700 and 1000 nm. Further, the lines for 700 and 1000 nm are nearly overlapping. This observation is consistent with the fluence distribution maps 
(Fig. \ref{McFluence}) and the PA signal profiles [see Fig. \ref{PAsignalSim}, panels (A)–(c), left panel]. It is clear that PA amplitude drops by nearly 35\% at 700 and 905 nm as the LL increases from 0 to 30\% and the same value is about 39\% at 1000 nm [see Fig. \ref{fig:SimD1_5_exp}(A)]. 
The estimated SO$_2$ values for PBS and PLS media are plotted in panels (G) and (g) of Fig. \ref{fig:SimD1_5_exp}, respectively, alongside the corresponding nominal LL values across the same wavelength combinations. For the 700-905 nm and 700-1000 nm wavelength pairs, the SO$_2$ estimates remain nearly constant with increasing LL. 
The corresponding PA signal profiles computed using the k-Wave toolbox for LL = 0\% and 30\% are shown in the left panel of Fig. \ref{PAsignalSim} (A)–(C) and (a)–(c), respectively. The PA signal strength at LL=0\% is greater than that of LL=30\%. Moreover, signal amplitude is consistently stronger at 905 nm than the other wavelengths. This is attributed to the higher values of $\mu_{\text{abs}}^{\text{BS}}$ at 905 nm for both LL=0\% and 30\% than those of the other wavelengths, as expected. 
The panels (A) and (a) of Fig. \ref{fig:SimD1_5_exp} display how $P_p$ varies with lysis level ranging from 0\% to 30\% for PBS and PLS suspending media, respectively at the optical wavelengths of 700, 905 and 1000 nm. The $P_p$ exhibits an almost linear decrease with increasing LL.
\clearpage
\newpage
\clearpage
%
\clearpage
\newpage
The maximum estimation errors in PBS are approximately 1.6\% and 3.6\% for these pairs, respectively. In PLS medium, the corresponding errors are 2.6\%  and 3.9\%, respectively. These values are tabulated in Tables \ref{700_1000exp} and \ref{700_905nm_exp} columns 4, 5 and 7, 8; rows 3-7. The SO$_2$ levels assessed using the 905-1000 nm combination have demonstrated significant fluctuations (data not shown). Logically, this combination is not suitable for the estimations of SO$_2$, H, and LL, as both wavelengths lie on the same side of the isosbestic point at 800 nm. Accordingly, results for this combination are not shown in Fig. \ref{fig:SimD1_5_exp}. 

The estimated H exhibits a linearly decreasing trend as shown in Fig. \ref{fig:SimD1_5_exp}, panels (M) and (m), for PBS and PLS media, respectively, across the same LL range and wavelength combinations. The highest estimation errors in PBS are approximately 8.2\%, and 10.9\% for the same wavelength pairs as mentioned earlier, respectively. In PLS medium, the respective errors are 10.2\%, and 8.4\%, respectively. In contrast, Fig. \ref{fig:SimD1_5_exp}, panels (S) and (s) reveal that evaluated LL rises linearly with nominal LL for PBS and PLS media, respectively. The estimation error increases as LL grows. In PBS, the average LL estimation errors are approximately 13.1\%, and 19\% for the 700-905 nm and 700-1000 nm combinations, respectively. Similarly, in PLS medium, the same errors are 18.1\% and 22.3\%, respectively. These values are detailed in Tables \ref{700_1000exp} and \ref{700_905nm_exp}, columns 2, 3 and 2, 6; rows 3-7.
%
%
%
\subsection{Experimental results}
Some representative experimentally obtained PA signals at 700, 905 and 1000 nm wavelengths and for LL= 0\% and 30\% are plotted in Fig. \ref{PAsignalSim} (D)-(F) and (d)-(f). The PA signal amplitude at LL=30\% is less compared to that of 0\%, as also seen in simulations (left panel of Fig. \ref{PAsignalSim}). In general, PA signal strength is higher at 905 nm than the other wavelengths. 

Figure \ref{fig:SimD1_5_exp}, panels (B)–(F) and (b)–(f), presents the impact of RBC lysis on peak-to-peak amplitude ($P_p$, mean $\pm$ std) for two different suspending media, at three different incident optical wavelengths and for blood samples collected from five different individuals (elaborated in Table \ref{donor}). The lysis level is gradually altered from LL=0 to 30\%. The experimental values of $P_p$ show a linear decrease with increasing LL. More specifically, approximately 35, 25 and 24\% reduction of PA amplitude can be seen at 700, 905 and 100 nm wavelengths for donor 1 [see Fig. \ref{fig:SimD1_5_exp}(B)]. 

The SO$_2$ levels estimated from experimental data do not exhibit any noticeable variation though LL has been changed from LL=0 to 30\% as shown in Fig. \ref{fig:SimD1_5_exp}(H). This trend is followed for both wavelength combinations and in both PBS and PLS suspending media [see Fig. \ref{fig:SimD1_5_exp}, panels (H)–(L) and (h)–(l)]. The computed values of blood SO$_2$ are also inserted in Tables \ref{700_1000exp} and \ref{700_905nm_exp}, columns 5 and 8, rows 8-32. In order to validate our findings, the same quantity has also been determined from the optical absorbance spectra measured using a UV-Vis spectrophotometer and the corresponding values are included in columns 4 and 7, Tables \ref{700_1000exp} and \ref{700_905nm_exp}. The average SO$_2$ estimation errors can be found to be around 14.8\%, 11.5\% for 700-905 nm and 10.2\%, 6.0\% for 700-1000 nm wavelength pairs in PBS and PLS media, respectively. The first combination clearly shows a little better accuracy than the later wavelength pair. Note that PA method seems to facilitate physiologically meaningful estimation of the blood SO$_2$ level since venous blood samples were collected from healthy individuals for which SO$_2$ level lies roughly between 60–80\%.
 
The assessed H decreases linearly as LL increases [see Fig. \ref{fig:SimD1_5_exp}, panels (N)-(R) and (n)-(r)]. It is clear from these figures that the PA method works very well up to LL$\le 10\%$. However, its deviation from the veridical line becomes pronounced at higher LL values, particularly at LL= 30\%. for the wavelength combinations 700-905 nm and 700-1000 and for both PBS and PLS media across all five donors. 

The estimated LL values also vary linearly with nominal LL levels as depicted in Fig. \ref{fig:SimD1_5_exp}, panels (T)-(X) and (t)-(x) and Tables \ref{700_1000exp} and \ref{700_905nm_exp}, columns 2, 3, 6 and rows 3-32. As expected, the mismatch between the evaluated and the actual values appears prominent at higher lysis levels, specifically at LL=30\%. In general, measured data for donor 5 demonstrate a larger variability compared to the remaining donors. 
%
%
\section{Discussion}
One of the primary contributions of this study is the adoption of the biconcave shape of RBCs, as opposed to the commonly assumed spherical geometry. The DDA method was employed to compute the absorption and scattering coefficient factors as well as the anisotropy factor for a single RBC. These parameters were evaluated for various surrounding media (PBS, PLS, PBS/PLS based lyzed media). Subsequently, optical parameters, namely, $\mu_{\text{abs}}^{\text{BS}}$, $\mu_{\text{sca}}^{\text{BS}}$ and $g$ were estimated for different blood samples.
\begin{table}[!h]
    \centering
    \caption{A comprehensive comparison of numerically and experimentally evaluated SO$_2$ values and lysis levels (LL) at the dual wavelengths of 700-905 nm.}
    \label{700_905nm_exp}
    \includegraphics[width=0.7\linewidth]{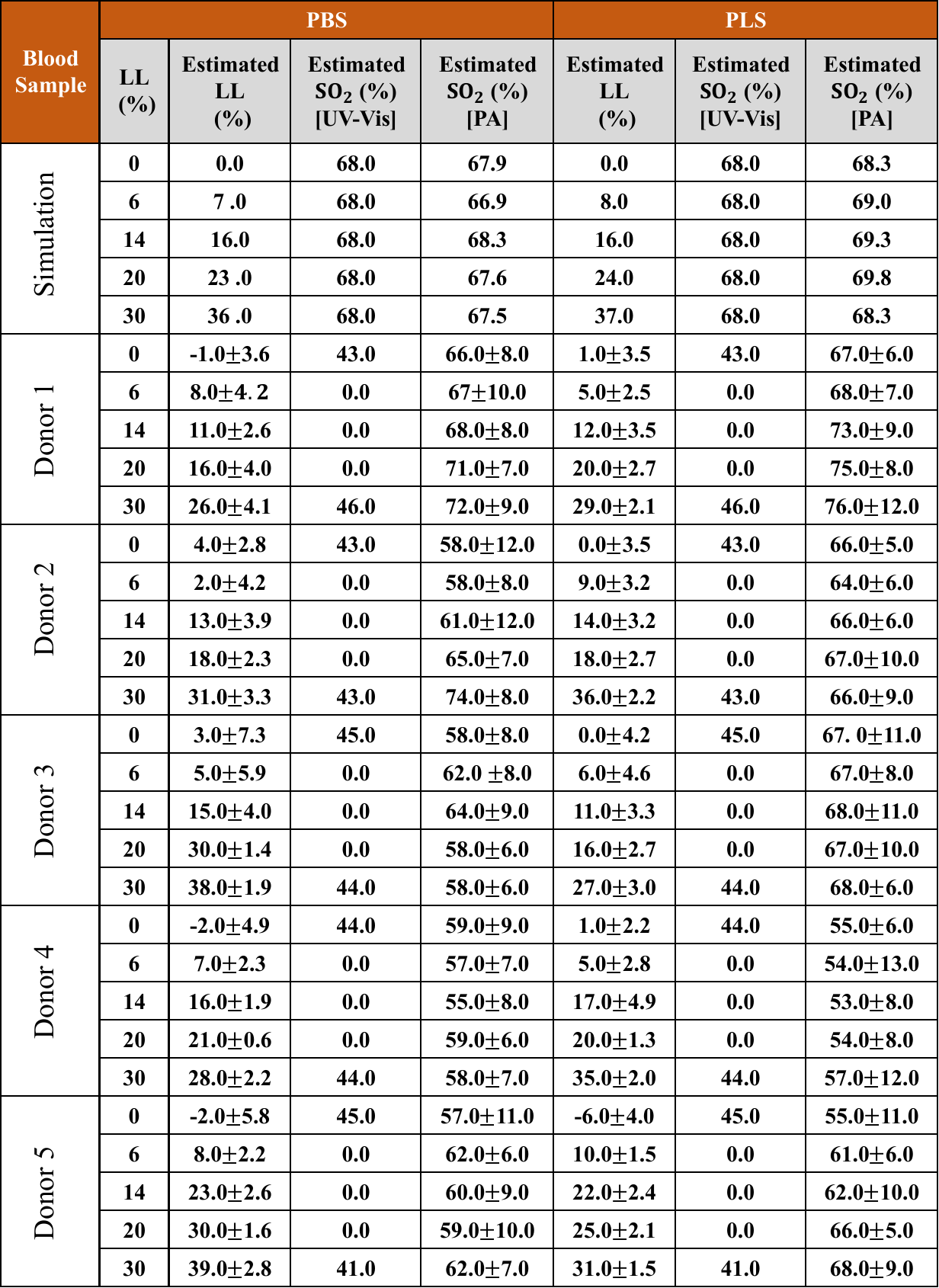}
\end{table}
\begin{table}[!h]
    \centering
    \caption{Detailed tabulation of the estimated SO$_2$ values and lysis levels (LL) determined via numerical and experimental investigations using the wavelength pair of 700-1000 nm.}
    \label{700_1000exp}
\includegraphics[width=0.7\linewidth]{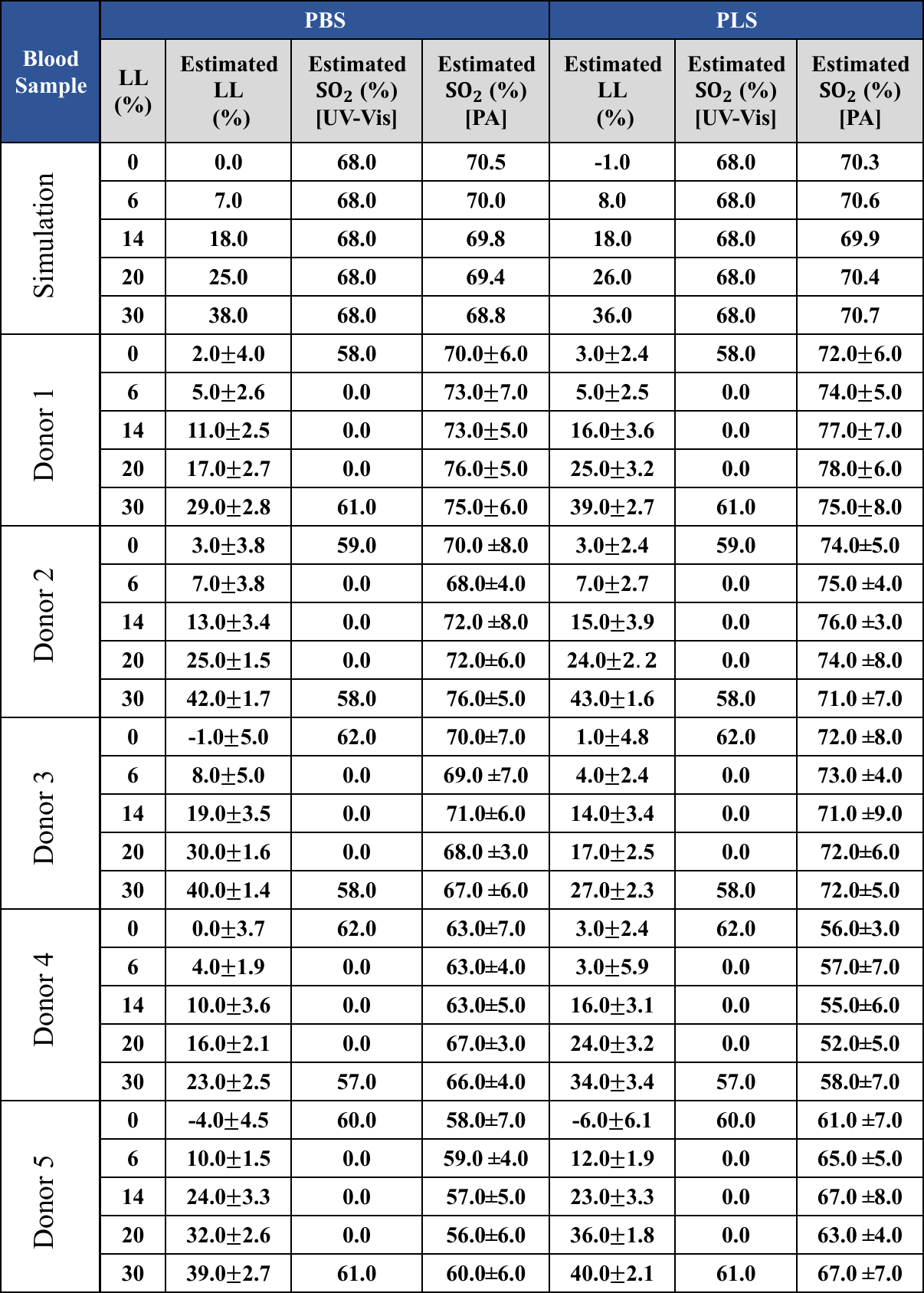}
\end{table}
 These parameters control photon propagation vis-\'{a}-vis PA emission from a tissue. DDA provides highly accurate results, but it is computationally intensive and time-consuming. For example, for a single RBC with 125 orientational average, computation time was approximately 3.3 hours at 1000 nm, whereas the time requirement was about 12 hours at 450 nm.
 Therefore, the computational challenge intensifies at lower wavelengths. It is worthy to point out that the number of dipoles used in this study was about $16\times 10^5$ at 1000 nm and $90\times 10^5$ at 450 nm. 
The DDA method may not be suitable to numerically quantify the optical parameters of blood samples containing clots or aggregates arising from multicellular interactions. Therefore, alternative methods are needed to efficiently estimate these parameters in such complex biological systems. MC simulations—especially GPU-accelerated versions like MCX—are widely used for estimating bulk optical properties in turbid media \cite{fang2009monte}. Finite-difference time-domain and finite element method approaches offer full-wave and multiphysics capabilities, making them well-suited for complex or heterogeneous tissues \cite{taflove2005computational}, \cite{jin2015finite}. These alternatives enable scalable and practical optical modeling beyond the scope of DDA. 
%

The numerical values of absorption coefficient, $\mu_{\text{abs}}^{\text{BS}}$, of the blood samples at different lysis levels considered in this study were calculated using Eq. (\ref{lyzed_blood_sample_Mua}) and are tabulated in Table \ref{nk_rbc_pbs_scott_Med}, columns 7 and 11, rows 3–17. The absorption values remain constant across all lysis levels at each wavelength. This is because light absorption depends on the total hemoglobin concentration, which does not change with lysis as the initial hematocrit remained fixed at H=50\%. The calculated values of $\mu_{\text{abs}}^{\text{BS}}$ provided by the DDSCAT algorithm look very interesting (see Table \ref{DDSCATopticalPRAM_pbs}, columns 4 and 9; rows 3-17). It decreases with increasing lysis level. This behavior essentially supports a fact that photons might have undergone multiple reflections inside erythrocytes (due to the presence of cell membrane) \cite{roggan1999optical}. Notably, as RBCs experience lysis, the sample became more homogeneous, leading to a reduction in the scattering coefficient ($\mu_{\text{sca}}^{\text{BS}}$) and an expected increase in the anisotropy factor ($g$). These variations can also be seen from Table \ref{DDSCATopticalPRAM_pbs}, columns 6, 7; 11, 12; rows 3-17). The changes of $\mu_{\text{abs}}^{\text{BS}}$, $\mu_{\text{sca}}^{\text{BS}}$ and PA amplitude with RBC lysis (columns 1 and 2 of Fig. \ref{fig:SimD1_5_exp}) are consistent with published results (see Fig. 10 of \cite{roggan1999optical}). 

It may be noted that RBCs undergo progressive biochemical and structural changes during storage, such as ATP depletion, increased membrane rigidity, oxidative stress, and ion imbalance. These changes make them more prone to lysis. Factors such as storage temperature, additive solutions, and leukoreduction significantly affect RBC viability. Additionally, exposure of phosphatidylserine and hemoglobin release due to lysis can trigger inflammation and transfusion-related complications, emphasizing the need for proper storage and monitoring of blood bags \cite{yoshida2019red,hess2010red}. The methodology reported herein may find an application to examine the health of stored blood. As found, 700-905 nm or 700-1000 nm wavelength pair may be utilized for simultaneous quantification of the oxygenation, hematocrit and lysis levels of such blood samples. Attempts will be made in future to achieve this end.

The technique may be further developed to quantify the lysis levels in vivo which may be useful for monitoring patients. Note that this method can accurately quantify the lysis level when the initial hematocrit is known. In practice, it is unlikely to have the knowledge of H before hand. This challenge may be solved using machine learning (ML), deep learning (DL) and artificial intelligence (AI) approaches, which may enable accurate and concurrent estimations of H, LL and SO$_2$ in vivo. 
\section{Conclusions}
The study investigates lysis process of RBCs using both numerical and experimental tools involving PAs. The DDA effectively captured the structural effects of RBCs, despite initial refractive index calculations did not incorporate cellular features. The estimation accuracy for H reached up to 90\% for 700-905\,nm and 700-1000\,nm wavelength pairs up to LL=14\%. Interestingly, SO$_2$ remained relatively stable across the LL range of 0-30\%. The nominal and predicted LL levels are found to be strongly correlated with correlation coefficient $\approx 90\%$. Among all tested combinations, the wavelength pairs 700-905 and 700-1000 nm have been found to be optimal for the simultaneous determination of H, LL and SO$_2$ parameters of blood samples in practice.
\subsection*{Acknowledgement}
SP thanks the members of the Biomedical Imaging Laboratory (BMIL), IIIT Allahabad for their continuous cooperation and support. Numerous discussions with laboratory members are also highly acknowledged. The authors are thankful to Mr. Vinod Tiwari, Mr. Anil Tiwari, Blood Bank, Swaroop Rani Nehru Hospital and Mr. Vinod Pathak, Health Center, IIIT Allahabad for providing blood samples. HSP acknowledges the support received from Dr. S. K. Majumder, Head, Laser Biomedical Applications Division, Raja Ramanna Centre for Advanced Technology, Indore (MP), India. The work was supported by the ICMR, DBT and ANRF grants (\# 56/2/2020-Hae/BMS, \# BT/PR44547/MED/32/791/2021 and \# CRG/2023/003278, respectively).
\subsection*{Conflicts of interests:}
The authors declare that there are no competing interests.
\subsection*{Data availability:}
All data needed to evaluate the conclusions in the paper are present in the paper and/or the Supplementary Materials. Additional data related to this paper may be requested from the authors.
\bibliography{library}
\bibliographystyle{abbrv}
\end{document}